\documentclass[aps, prd, twocolumn, amsmath, floats,floatfix, superscriptaddress, nofootinbib,
showpacs]{revtex4}

\usepackage{amssymb}
\usepackage{amsmath}
\usepackage{verbatim}
\usepackage{mathrsfs}
\usepackage{amsfonts}
\usepackage{latexsym}
\usepackage{epsfig}
\usepackage{color}
\usepackage{graphicx,subfigure}
\usepackage{units}
\usepackage{slashbox}

\begin{document}

%%%%%%%%%%%%%%%%%
%%%   TITLE   %%%
%%%%%%%%%%%%%%%%%
\title{Time evolution of superradiant instabilities for charged black holes in a cavity}  
% Othre titles suggested:

 \author{Juan~Carlos~Degollado}\email{jcdaza@ua.pt}
 \author{Carlos~A.~R.~Herdeiro}\email{herdeiro@ua.pt}
   \affiliation{
   Departamento de F\'\i sica da Universidade de Aveiro and I3N, 
   Campus de Santiago, 3810-183 Aveiro, Portugal.
 }

%%%%%%%%%%%%%%%%
%%%   DATE   %%%
%%%%%%%%%%%%%%%%

\date{December 2013}

%%%%%%%%%%%%%%%%%%%%
%%%   ABSTRACT   %%%
%%%%%%%%%%%%%%%%%%%%

\begin{abstract}  
Frequency domain studies have recently demonstrated that charged scalar fields exhibit fast growing
superradiant instabilities when interacting with charged black holes in a cavity. Here, we present a
time domain analysis of the long time evolution of test charged scalar field configurations on the
Reissner-Nordstr\"om background, with or without a mirror-like boundary condition. Initial data is taken
to be either a Gaussian wave packet or a regularised (near the horizon) quasi-bound state. Then,
Fourier transforming the data obtained in the evolution confirms the results obtained in the
frequency domain analysis, in particular for the fast growing modes. We show that spherically
symmetric ($\ell=0$) modes have even faster growth rates than the $\ell=1$ modes for `small' field charge. Thus, our study
confirms that this setup is particularly promising for considering the non-linear development of the
superradiant instability, since the fast growth makes the signal overcome the numerical error that
dominates for small growth rates, and the analysis may be completely done in spherical symmetry.
 
%since: 
%1) the fast growth rates avoid the contamination by numerical errors that
%can occur for small growth rates; 
%2) the analysis may be completely done in spherical symmetry; 
%3) the mirror-like boundary conditions
%are under control for a time evolution.

\end{abstract}

%%%%%%%%%%%%%%%%
%%%   PACS   %%%
%%%%%%%%%%%%%%%%


%%%%%%%%%%%%%%%%%%%%%%
%%%   MAKE TITLE   %%%
%%%%%%%%%%%%%%%%%%%%%%

\maketitle

%%%%%%%%%%%%%%%%%%%%%%%%%%%%%%%%%%%%%%%%%%%%%%%%%%
\section{Introduction}\label{sec:introduction}
%%%%%%%%%%%%%%%%%%%%%%%%%%%%%%%%%%%%%%%%%%%%%%%%%%
Superradiant scattering is a classical process through which energy can be extracted from a black
hole (BH). For a rotating, i.e Kerr, BH,  with horizon angular velocity $\Omega_+$, this
process occurs when an impinging wave of a bosonic field with frequency $\omega$ and azimuthal quantum number $m$ obeys
the condition $\omega<m\Omega_+$
\cite{Bardeen:1972fi,Starobinsky:1973a,Press:1972zz,Zouros:1979iw,Cardoso:2004nk,Dolan:2007mj}.
Then, upon
scattering, rotational energy is extracted from the BH, simultaneously extracting enough angular
momentum such that the BH's area/entropy does not decrease. This scattering can become an
instability if the wave repeatedly scatters off the BH. This is achieved, for instance, by either
introducing a mass term for the field or imposing a mirror-like boundary condition outside the BH,
which reflects the wave. Then, the total extracted energy grows exponentially with time - suggesting
an explosive phenomenon that has been dubbed the \textit{black hole bomb}
\cite{Press:1972zz,Cardoso:2013krh}  -,
and the linear (test field) approximation will 
break down. The complete non-linear development of this instability, including its back-reaction on
the geometry and its endpoint, is still not understood. It has been suggested that the explosive event
may be akin to a \textit{bosenova} \cite{Yoshino:2012kn} observed in condensed matter systems, and
further progress needs fully non-linear numerical simulations \cite{Cardoso:2012qm}. One problem
in achieving these numerical  evolutions  is that the growth rate of the instability is very small
\cite{Cardoso:2004nk,Berti:2009kk,Hod:2009cp,Rosa:2009ei,Pani:2012vp,Witek:2012tr,Dolan:2012yt} and
its signal can become buried in numerical noise.

In a recent work \cite{Herdeiro:2013pia} we have shown the growth rate of scalar superradiant modes can 
be much
faster in the case of a charged, i.e  Reissner-Nordstr\"om (RN), BH. In this case, 
superradiant scattering occurs if a charged field with frequency $\omega$ and charge $q$ impinging
on a charged BH with horizon electric potential $\Phi_+$ obeys
$\omega<\omega_c\equiv q\Phi_+$
\cite{Bekenstein:1973mi}. Upon scattering, Coulomb energy is extracted from the BH, 
simultaneously decreasing enough its charge such that its area/entropy does not decrease. In this
case, however, in order to have a superradiant instability it is
not enough to add a mass term to the field. It is also necessary to impose a mirror-like boundary
condition at some distance from the BH
\cite{Furuhashi:2004jk,Hod:2012zz,Hod:2013nn,Konoplya:2013rxa}. For such charged BH in a
cavity, states with an 
imaginary part of the frequency of up
to Im$(M\omega)\sim 0.07$ were obtained in \cite{Herdeiro:2013pia}. As a comparison, in the Kerr
case, the maximum growth rates are Im$(M\omega)\sim  1.74\times 10^{-7}$ for massive fields in 
asymptotically Minkowski space and 
Im$(M\omega) \sim 6\times 10^{-5}$ for mirror-like boundary
conditions \cite{Cardoso:2004nk,Dolan:2012yt}. The growth time scale is set by the inverse of
Im$(\omega)$. Therefore the aforementioned frequencies demonstrate that the e-folding time can 
be at least three orders of magnitude smaller in
the charged case than in the rotating case.
Subsequently, it was argued in \cite{Hod:2013fvl} that, for the charged case, Im$(M\omega)$ scales
as
$qQ/r_+$, for large $qQ$, where $Q,r_+$ are the charge and (Schwarzschild) radial coordinate of the
BH's event horizon. Such scaling implies that in the charged case, the instability's e-folding time
can be made arbitrarily short by increasing $q$.

In this paper, by
performing a time domain analysis, we further confirm the existence of fast growing superradiant
modes for charged BHs in a cavity. We evolve in time two types of charged scalar field wave packets,
in
the background of a RN BH in a cavity with radius $r_m$. The choice of the initial data,
as well as background and field parameters determines the particular set of quasi-bound states that will have leading amplitudes. By time
evolving this initial data and Fourier transforming the evolution data, we identify the frequencies of these dominant quasi-bound
states, and observe they match values obtained in the frequency domain analysis \cite{Herdeiro:2013pia}, some of which correspond
to unstable modes. After a sufficiently long time evolution, we verify that the unstable mode with the
largest growth rate dominates the evolution and its growth rate is in excellent agreement with the imaginary part
obtained from the frequency domain analysis. 

In  \cite{Herdeiro:2013pia}  we have focused on modes with angular momentum number $\ell=1$. This
choice was motivated by the goal of comparison with the rotating case, for which these modes are the
fastest growing ones. Here, we shall also explicitly consider $\ell=0$ modes and observe that
similar instabilities to the $\ell=1$ case -  which may have even faster growth rates for `small'  field charge - are obtained. By contrast, in the 
rotating case, modes with $\ell=0$ cannot yield superradiant states. The existence of unstable
spherically symmetric modes for the charged case  implies that the non-linear development of the
superradiant instability can be studied in a spherically symmetric setup. Such symmetry, together
with the shorter time scales involved, makes charged BHs in a cavity particularly interesting for non-linear
studies of the superradiant instability. 

One may worry, however, that numerical difficulties will be introduced by the need to impose a
mirror-like boundary condition. Indeed, in \cite{Witek:2010qc}, for instance, an attempt to impose
such boundary conditions for a binary BH evolution led to loss of convergence after a few
reflections
of the signal off the reflecting boundary. We will find no such problem:  our 
implementation of the mirror-like boundary condition yields long term stable evolutions, as in
\cite{Dolan:2012yt}, for the test field analysis performed herein.

This paper is organized as follows. In Sec. \ref{sec:frame} we describe the background and field setup.
In Sec. \ref{tdomain} we discuss the test field's flux and energy which shall be monitored during the time evolutions, 
 the system of equations to be solved in the time domain analysis, the boundary conditions and the numerical method.  In Sec. \ref{results}  the numerical results
for two types of initial data are described.  We also review some results on quasi-bound 
states of charged scalar fields around charged BHs with or without mirror-like boundary conditions. In the following, the former states shall be dubbed \textit{mirrored states}. Final remarks are presented in Sec. \ref{conclusions}. 
We shall use units in which $G=1=c=\hbar$.

%Superradiant instability of a charged scalar field  in a RN background happens, when the 
%black hole is surrounded by a mirror that scatters the superradiant waves back to the horizon,
%amplifying it at each scattering. It has been shown that the total extracted energy grows
%exponentially with time.

%To investigate the dynamics of unstable modes we consider the temporal evolution
%of a charged scalar field with an unstable mode in the Reissner Nordstrom geometry. 

%The growth rate of the instability is very small and it is expected the effect of the
%instability will appear in the late time behavior. However, it becomes difficult to detect the
%instability numerically because the signal can be buried by numerical noise.
%Some progress have been done in this direction for a massive scalar field in the background of a
%Kerr black hole \cite{Dolan:2012yt,Witek:2012tr}.

%%%%%%%%%%%%%%%%%%%%%%%%%%%%%%%%%%%%%%%%%%%%%%%%%%
\section{Framework}\label{sec:frame}
%%%%%%%%%%%%%%%%%%%%%%%%%%%%%%%%%%%%%%%%%%%%%%%%%%
We consider Einstein-Maxwell theory minimally coupled to a complex, charged, massive scalar field.
Linearizing the field equations (on the scalar field), yield the Einstein-Maxwell equations plus the
Klein-Gordon equation describing the evolution of massive charged
scalar perturbation in the Einstein-Maxwell background:
\begin{equation}\label{eq:KG}
 (\nabla_{\nu}-iqA_{\nu})(\nabla^{\nu}-iqA^{\nu})\Phi-\mu^2\Phi = 0 \ ,
\end{equation}
where $A_{\nu}$ is the electromagnetic potential and $q$ and $\mu$ are the charge and the mass of
the scalar field.

%What is really new: long time evolutions 

The Einstein-Maxwell background we wish to consider is the RN black hole. The metric, written in
standard Schwarzschild type coordinates $(\tilde{t},r,\theta, \phi)$ reads
\begin{equation}
ds^2=-f(r)d\tilde t^2+\frac{dr^2}{f(r)}+r^2(d\theta^2+\sin^2\theta d\phi^2) \ ,  
\end{equation}
where 
\begin{equation}
f(r)=\frac{(r-r_+)(r-r_-)}{r^2} \ , \ \ r_\pm \equiv M\pm\sqrt{M^2-Q^2} \ ,
\end{equation}
whereas the gauge field is given by an electric potential of the form $A_\nu
dx^\nu=-Q/r \, d\tilde t$.

The Klein-Gordon equation \eqref{eq:KG} is separable in the RN geometry
with the ansatz
\begin{equation}\label{eq:scalar_decomp}
 \Phi(\tilde t,r,\theta,\varphi) = \sum_{\ell,m} e^{-i\omega \tilde t}Y_{\ell}
^m(\theta, \varphi) \phi_{\ell}(r) \ ,
\end{equation}
where $ Y_{\ell}^m(\theta, \varphi)$ are the scalar spherical harmonics satisfying $\Delta_{S^2}
Y_{\ell}^
m(\theta, \varphi)=-\ell(\ell+1)Y_{\ell}^m(\theta, \varphi) $. A Schr\"odinger-like radial equation
equation is then obtained for 
a new radial function
$R_{\ell}(r)=r\phi_{\ell}(r)$:
\begin{equation}
\label{eq:Slike}
\left[-\frac{d^2}{dr^{*2}}+V(r)\right]R_{\ell}(r)=\omega^2R_{\ell}(r) \ , 
\end{equation}
where the effective potential is given by 
\begin{equation}
 \label{eq:Veff}
V(r)= \frac{2qQ\omega}{r}-\frac{q^2Q^2}{r^2}+f(r)\left(
\frac{l(l+1)}{r^2}+\mu^2+\frac{f'(r)}{r} \right) \ .
\end{equation}
The tortoise coordinate $r^*$ is defined as 
\begin{eqnarray}
 r^* &\equiv &\int \frac{1}{f(r)} dr  \\
 &=& r +\frac{r_{+}^2}{r_{+}-r_{-}}\ln (r-r_{+})- \frac{r_{-}^2}{r_{+}-r_{-}}\ln (r-r_{-}) \ ,
\nonumber 
\end{eqnarray}
The quasi-bound state solutions we wish to consider satisfy a purely ingoing boundary condition 
close to the external horizon, $R_{\ell}\sim e^{-i(\omega-\omega_c)r^{*}}$. In the far field region
they should decay to zero; asymptotically these solutions can be written as
$R_{\ell}\sim e^{ \tilde q r}$ with $\tilde q=\sqrt{\mu^2-\omega^2}$, and  ${\rm Re}(\tilde q) <0$.
As a consequence 
of imposing a pair of boundary conditions the spectrum of complex frequencies
becomes
discrete. The real part of $\omega$ sets the oscillation frequency of the
mode, whereas the imaginary part determines its growth or decay rate.

%%%%%%%%%%%%%%%%%%%%%%%%%%%%%
\section{Time domain analysis}\label{tdomain}
%%%%%%%%%%%%%%%%%%%%%%%%%%%%%
%Here we solve the scalar wave equation in a fixed RN background in Kerr-Schild coordinates using a
%1+1 numerical evolution code.
In order to solve the Klein-Gordon equation in the time domain, 
we use a more suitable coordinate system, the ingoing Kerr-Schild coordinates
$(t,r,\theta,\phi)$, obtained from Schwarzschild coordinates by changing the time coordinate 
\begin{equation}
 t= \tilde t + (r^* - r) \ .
\end{equation}
In these coordinates the RN metric becomes
\begin{eqnarray}
& ds^2_{KS} =-\left(1-\frac{2M}{r} + \frac{Q^2}{r^2}\right)d{t}^2 + 2\left(\frac{2M}{r}-
\frac{Q^2}{r^2}\right)dtdr &\nonumber\\ 
&+\left(1+\frac{2M}{r}-\frac{Q^2}{r^2}\right)dr^2 + r^2(d\theta^2+\sin^2\theta d\phi^2) \ . & 
\end{eqnarray}
In terms of the standard ADM variables, $ds^{2}=-(\alpha ^{2}-\beta ^{i}\beta _{i})dt^{2}+2\beta
_{i}dx^{i}dt+\gamma_{ij}dx^{i}dx^{j}$, we read off that
% the metric is written as
%\begin{equation}
%ds^{2}=-(\alpha ^{2}-\beta ^{i}\beta _{i})dt^{2}+2\beta _{i}dx^{i}dt+\gamma
%_{ij}dx^{i}dx^{j}.
%\end{equation}
%Where 
%\begin{equation}
%ds^2=-(\alpha^2-\beta_{r}\beta^{r})dt^2 +2\beta_{r} dtdr+\gamma_{rr}dr^2+r^2d\Omega^2\,
%\end{equation}
the lapse function $\alpha$ and the shift vector $\beta^{i}$ are 
\begin{eqnarray}
 \alpha=\left(1+\frac{2M}{r}-\frac{Q^2}{r^2} \right)^{-1/2} \ , \quad \beta_{r} =
\frac{2M}{r}-\frac{Q^2}{r^2}\ .
\end{eqnarray}
Observe that $\beta^{i}=[\beta_{r}\alpha^2,0,0]$.
Furthermore, the radial components of the induced metric $\gamma _{ij}$ are
\begin{equation}
\gamma_{rr}= \frac{1}{\alpha^2},\quad \gamma_{\theta\theta}=r^2 ,\quad
\gamma_{\varphi\varphi}=r^2\sin^2{\theta}\ .
\end{equation}
%

%In these coordinates radially ingoing null lines have constant coordinate velocity $dr/dt=-1$. The
%outgoing null lines, have coordinate velocity given by $ -(r-r_{+})(r-r_{-}) /(r^2+2M-Q^2)$; in other
%words, their speed is less than 1 for $r>r_{+}$, it is zero for $r=r_{+}$ and becomes negative for
%$r_{-}<r<r_{+}$. Since all the outgoing lines go in, this region can have no causal influence
%on the outside. {\bf unclear!!!} 

The Kerr-Schild form of the metric has become popular in the numerical relativity
community because a constant $t$ hypersurface is non singular and horizon penetrating which allows
for convenient imposition of boundary conditions.

%%%%%%%%%%%%%%%%%%%%
\subsection{Scalar flux and energy}
%%%%%%%%%%%%%%%%%%%%

In the RN spacetime the timelike Killing vector field $k=\partial/\partial t$ defines a conserved
current $J^{\mu}=T^{\mu}{}_{\nu}k^{\nu}$. 
The integration of the conservation law $\nabla_{\mu}J^{\mu}=0$ over a finite region of
spacetime yields
\begin{equation}
 \int \nabla_{\mu}J^{\mu}\, \sqrt{-g}\, d^{4}x  =0\ .
\end{equation}
Since the spacetime is static, the last integral becomes an integration over a surface of
$t={\rm cte}$, $\Sigma_t$, 
\begin{equation}
 \int_{\Sigma_t}  \nabla_{\mu}J^{\mu}\,\alpha\sqrt{\gamma}\, d^{3}x=0\ ,
 \label{cons}
\end{equation}
in terms of the standard ADM variables $\alpha$ and $\gamma$ (determinant of the spatial metric). 
Taking the field energy $E$ as the integral on a space-like slice of the zeroth component of the
currrent
\begin{equation}\label{eq:ener}
 E= \int_{\Sigma_t}  \alpha\,J^{0} \, \sqrt{\gamma}\,d^3x \ , 
\end{equation}
then (\ref{cons}) implies
\begin{equation}
 \frac{d}{dt}E =-\int_{\Sigma_t}d^{3}x\partial_{i}(\alpha \sqrt{\gamma}J^{i}) = -\int_{\partial
\Sigma_{t}}\alpha J^{i}\ dS_{i} \ ,
\end{equation}
where $\partial \Sigma_{t}$ is the boundary of the hypersurface,  $dS_{i}=n_{i}
d\theta
d\varphi \sqrt{g_{_{S^2}}}$ and $n^{i}$ is the vector normal.

Using the decomposition in spherical
harmonics of the field \eqref{eq:scalar_decomp} the energy \eqref{eq:ener},  can be written as
\mbox{$E=\sum\limits_{\ell, m} E_{\ell m}$}, with
\begin{equation}\label{eq:energy}
E_{\ell m} = \int_{r_{+}}^\infty \rho_{\ell}(r) r^2\,dr\; , 
\end{equation}
where
\begin{eqnarray}
\rho_{\ell }(r) &=&  \displaystyle{\frac{1}{2}\left[\frac{1}{\alpha^2} \left(
|\partial_{t}\phi_{\ell }|^2-2 qA_{t}{\rm Im}(\phi^{*}_{\ell}\partial_{t}\phi_{
\ell})  \right) +(1-\beta_r)\times \right.} \nonumber \\ 
&& \ \ \ \displaystyle{\left.  \times |\partial_{r}\phi_{\ell}|^2+ \left(
\frac{q^2A_{t}^2}{\alpha^2}+
\frac{\ell(\ell+1)}{r^2}+\mu^2\right)|\phi_{\ell }|^2 \right]} \ . \nonumber
\end{eqnarray}%

In a similar way, the flux passing though a surface of $r={\rm cte}$
\begin{equation}
F^{r}\equiv  -\int_{\partial
\Sigma_{t}}\alpha J^{i}\ dS_{i} = -\int_{\partial\Sigma_t} T^{r}_{\ 0} r^2\, d\theta d\varphi
\sin\theta \ ,
\end{equation}
can be decomposed as a sum over angular modes, 
\mbox{$F^{r}=-\sum\limits_{\ell, m}r^2\, F^{r}_{\ell m}$},
where
%\begin{equation}
% F^{r}_{lm} =-\int \rho^{c}_{lm} r^2 \sin{\theta}d\theta d \varphi = -\rho^{c}_{lm} r^2 \ ,  
%\end{equation}
\begin{eqnarray*}
F^{r}_{lm} & = & \displaystyle{(1-\beta_r)\left[{\rm Re}
(\partial_{r}\phi_{\ell}\partial_t\phi_{\ell}^{*}) - qA_{t}{\rm
Im}(\phi_{\ell}^{*}\partial_r\phi_{\ell})\right]} \\ \nonumber
&&\displaystyle{+ \beta_{r}\left[|\partial_{t}\phi_{\ell}|^2
-2qA_{t}{\rm Im }(\phi_{\ell}^{*}\partial_{t}\phi_{\ell})+q^2A_{t} \nonumber
^2|\phi_{\ell}|^2\right ] \ .}
\end{eqnarray*}

\subsection{Equations of motion}

To obtain a first order system of equations of motion, we introduce the auxiliary functions
\begin{equation}
\psi=\partial_r\phi \ , \qquad  
\pi=\frac{1}{\alpha^2}(\partial_t\phi -\beta^r\psi) \ ,
\end{equation}
where we have dropped the mode number subscripts to
simplify
the notation. The following system of evolution
equations for $\phi$, $\pi$ and $\psi$ is obtained.
\begin{equation}\label{eq:phi_t}
 \partial_{t}\phi=\alpha^2( \pi + \beta_{r}\psi)\ ,
\end{equation}
\begin{eqnarray}\label{eq:psi_t}
 \partial_{t}\psi&=&\partial_{r}( \alpha^2 (\pi + \beta_{r}\psi ) )\\ \nonumber
&=&\beta^r\partial_{r}\psi +  \alpha^2\partial_{r}\pi+
\frac{2r(Mr-Q^2)}{(r^2+2Mr-Q^2)^2}(\pi-\psi)\ ,
\end{eqnarray}
\begin{eqnarray}\label{eq:pi_t}
 \partial_{t}\pi&=&\beta^{r}\partial_{r}\pi+\alpha^2\partial_{r}\psi\\ \nonumber
&&-\left(\frac{\ell(\ell+1)}{r^2} -\frac{q^2Q^2}{\alpha^2r^2}
+\mu^2\right)\phi-\frac{2iqQ}{r}\pi+\frac{iqQ^3}{r^4}\phi\\
\nonumber
&&+\frac{2}{(r^2+2Mr-Q^2)^2}( T_1\pi+ T_2\psi  ) \ ,
\end{eqnarray}
where
\begin{eqnarray*}
&T_1\equiv \frac{1}{r}( Mr^3+4M^2r^2-4MrQ^2+Q^4 )\ , &\\
&T_2\equiv r(r^2+3Mr-2Q^2) \ .&
\end{eqnarray*}

%The  first order variables conveniently allows us to write these equations as a linear system of the
%form
%\begin{equation}
% \partial_t {\bf u} -{\bf A}\partial_{r} {\bf u} =s({\bf u})
%\end{equation}
%where the state vector is
%\begin{equation}
% {\bf u} =\left(\pi,\psi\right)^{T}
%\end{equation}
%and 
%\begin{equation}
%{\bf A}=\left(
%\begin{array}{cc}
%\beta ^{r} & \alpha ^{2} \\
%\alpha ^{2} & \beta^{r}
%\end{array}
%\right) .
%\end{equation}

%We have not considered the equation for $\phi$ since it can be recovered from the definition of 
%$\pi$. From this form we can perform a linear analysis. Initially the scalar field will propagate
%as a wave with the moving -out and moving -in components.

%The system is symmetric hyperbolic, so the quantities $\phi$, $\psi$ and $\pi$ may be decomposed
%into characteristic fields that propagate with well determined characteristic speeds. Although the
%characteristic fields propagate causally, due to the mass term the relation of dispersion is
%modified and the group and phase velocity do not coincide anymore.

The definition of $\psi$ becomes a constraint, of the form
\begin{equation}
 C_r(r,t) = \partial_r\phi(r,t) -\psi(r,t) \ ,
\end{equation}
that must be satisfied at all times. If $C_r$ is zero initially, and the evolution equation holds
exactly, $C_r$ will remain zero during the evolution. Numerical truncation errors and
boundary errors may cause, however, deviations from zero. Hence, keeping track of the
evolution of $C_r$ provides a test for the accuracy of the numerical simulations.

\subsection{Numerics and boundary conditions}

The evolution equations for the radial components were solved
by making use of the 1+1 dimensional PDE solver described in \cite{Degollado:2009rw} and also used
in \cite{Burt:2011pv}. 
The numerical algorithm utilized by this
code is based on the method of lines in a third order Runge-Kutta scheme such that
the spatial derivatives were evaluated with a second order symmetric finite difference
stencil. To guarantee stability by suppressing high frequency instabilities a standard
fourth order dissipation term was also applied.
% in solving the evolution equations for the
%multipole components

We have decomposed the field into its real and imaginary part $\phi=\phi^{R}+i\phi^{I}$ and 
solved the corresponding system of coupled equations.
%
%The inner radius was chosen to be Rin = 1.3M and the outer boundary was chosen
%to be at R_out = 4000M
The evolution scheme is simpler than for the corresponding Kerr case  
\cite{Strafuss:2004qc,Witek:2012tr}; but it is still technically
challenging because of the requirement of high resolution, long integration times and boundary 
conditions.

A simple way to implement the mirror-like boundary condition at a given radial coordinate 
$r=r_m$ is by setting the field and the dynamical fields to
zero at the boundary. This works well when the shift vector is equal to zero. For the RN spacetime in Kerr-Schild
coordinates, however, this boundary condition produces strong violations of the constraint. These constraint
violations propagate into the inner domain and grow, eventually dominating the numerical solution.

In order to impose the mirror-like boundary condition in our numerical scheme we proceed as follows.
At the mirror position the field must satisfy three conditions $\phi(r_m)=0$, $\partial_t \phi
(r_m)=0$ and $\partial_{tt} \phi (r_m)=0$. The first condition is straightforward to implement. The
second, using equation \eqref{eq:phi_t} gives a relationship between $\psi (r_m)$ and $\pi(r_m)$.
For the third we substitute the definition of $\pi$,
$\partial_{tt}\phi(r_m)=\partial_{t}(\alpha^2\pi(r_m)+\beta^{r}
\psi(r_m))=\alpha^2\partial_t\pi(r_m)+\beta^r\partial_t\psi(r_m)$ and use the equation
\eqref{eq:pi_t} together with \eqref{eq:psi_t}
to get and equation for $\pi(r_m)$, $\psi(r_m)$ and their spatial derivatives. Finally we use one
sided finite differencing stencils to approximate the radial derivatives. This procedure provide us
an
algebraic system of equations for $\psi(r_m)$ and $\pi(r_m)$ in terms of their values at the inner
points that has to be solved at each time step.

%%%%%%%%%%%%%%%%%%%%%%%
\section{Results}
\label{results}
%%%%%%%%%%%%%%%%%%%%%%%
Solving the above system of equations of motion one obtains a time series for the scalar field
amplitude at an observation point with a given radius $r_o$, with $r_+<r_o<r_m$. After a given
number of time steps we perform a Fast Fourier transform to obtain the field amplitude in frequency
space - the power spectrum - and identify the frequencies. In the following we
shall report two distinct cases. 
The evolution of the charged scalar field in 1) the asymptotically Minkowski RN BH; 2) in the RN BH
in a cavity.

\subsection{Asymptotically Minkowski space}

%In this section we present the frequency spectra determined with the continued fraction approach
%o validate the results numerically obtained 

%The Quasi bound state excitation depends on the structure of the scalar field that is used
%as a perturbation. To excite certain modes in a controlled way, we choose initial data of the form
%of a Gaussian.

We have evolved the charged scalar field in the RN background 
\textit{without} imposing the mirror-like boundary condition. Instead, we impose the condition $\pi+\psi =0$ at some $r=r_{max}$. For
an uncharged massless scalar this corresponds to impose the incoming characteristic modes to be zero.
For a charged massive scalar field this condition introduces a small error that propagates inside
the numerical domain. In order to avoid any kind of contamination we set $r_{max}$ sufficiently far out, i.e $r_{max} > t_{evol}$ where $t_{evol}$ is the total evolution time.

\begin{widetext}

\begin{figure}[h!]
\begin{center}
\includegraphics[width=0.37\textwidth,clip,angle=-90]{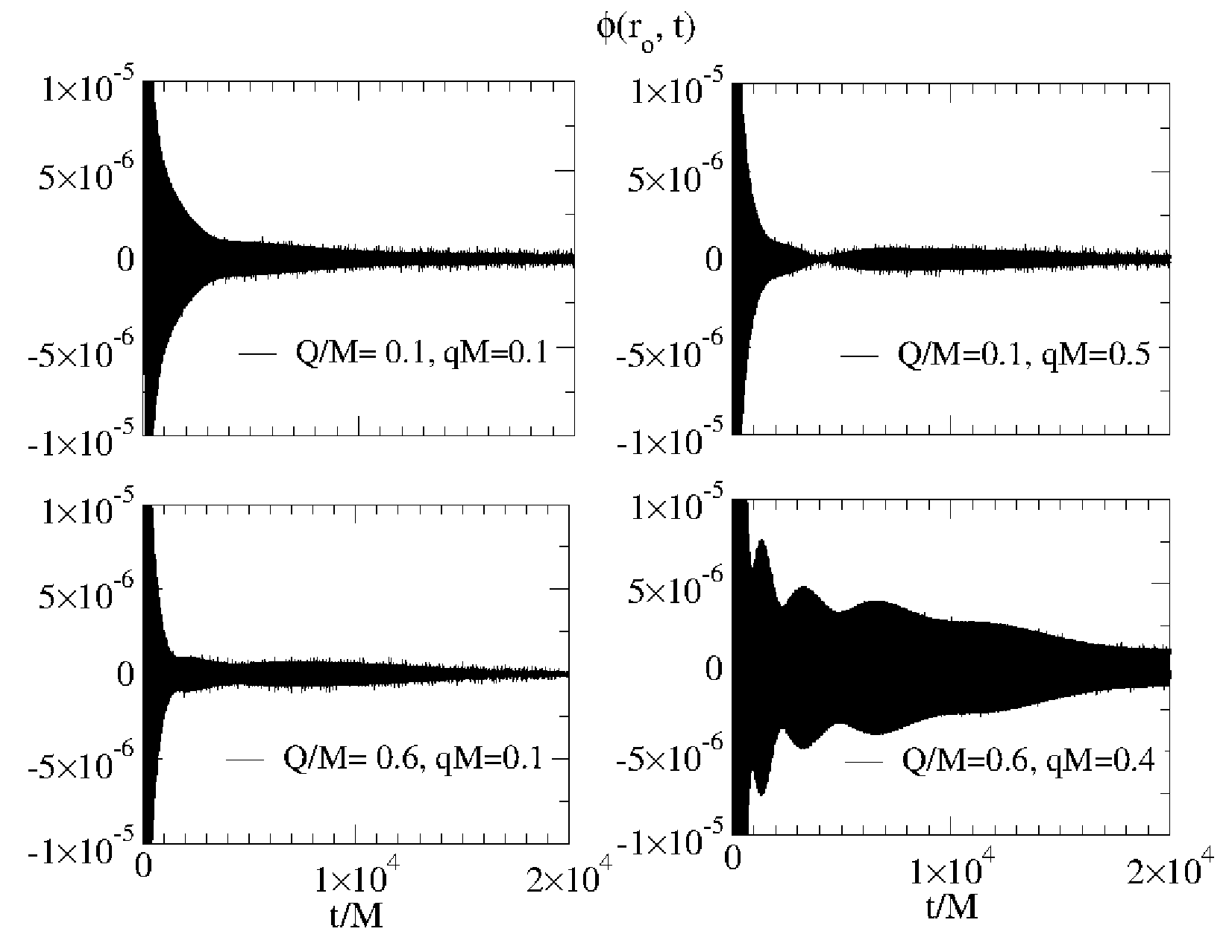}
\includegraphics[width=0.37\textwidth,clip,angle=-90]{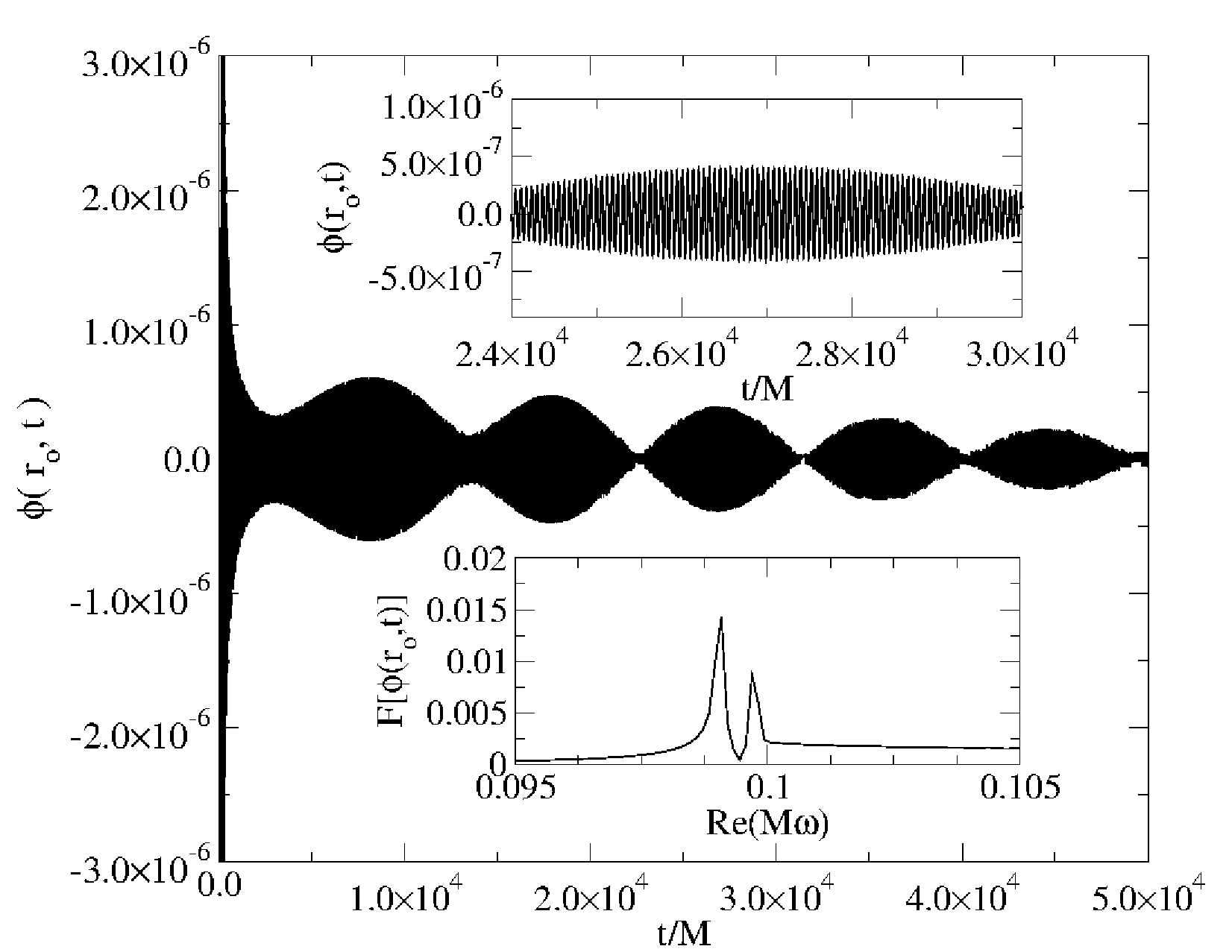}
\caption{Evolution of the scalar amplitude for a spherical ($\ell=0$) Gaussian perturbation with $\phi_0=3\times 10^{-3}$, $r_g=10M$, $\sigma =2M$. (Left panel) Decaying patterns for $r_0 = 10M$, $M\mu=0.3$ and various values of $Q,q$. $M$ was set to one in the figure. (Right panel) Beating pattern obtained for $r_0 = 70M$, $M\mu=0.1$, $Q/M=0.1$ and $Mq=0.3$. The upper inset is a zoom showing the oscillating behaviour; the Fourier transform is shown in the bottom inset and exhibits the two excited modes. 
}
\label{fig:mu03_l0_obsR1} 
\end{center}
\end{figure}

\end{widetext}

 Fig. \ref{fig:mu03_l0_obsR1} (left panel)
illustrates a generic evolution. We have used as  initial data at $t=0$ a static Gaussian
perturbation centered at $r_{cg}$:
\begin{equation}
 \phi(r,0)=\phi_0 e^{-(r-r_{cg})^2/2\sigma^2 } \ , \quad \partial_{t}\phi(r,0) = 0 \ , 
 \label{gaussian}
\end{equation}
with $\phi_0=3\times10^{-3}$, $r_{cg}=10M$
and $\sigma=2M$. The power spectrum obtained from this time evolution data,
shows set of distinct frequencies that can be compared with those obtained in the frequency
domain in \cite{Degollado:2013eqa,Herdeiro:2013pia} for the same parameters. This comparison 
is made in Table \ref{tab:num} and shows a good agreement.

\begin{center}
\begin{table}[h!]
  \begin{tabular}{ |c |c | c | c| c|}
    \hline
$Q/M$ & $qM$ & ${\rm Re}(M\omega)_{Analytical}$ & ${\rm Re}(M\omega)_{Numerical}$  \\ 
\hline \hline
$0.1$ & $0.1$ &$0.2875$ & $0.287$ \\
\hline
$0.1$ & $0.5$ &$0.2900$ & $0.290$ \\
\hline
$0.6$ & $0.1$ &$0.2903$ & $0.290$ \\
\hline
$0.6$ & $0.4$ &$0.2993$ & $0.299$ \\
\hline
  \end{tabular}
\caption{Comparison between analytic and numerical frequencies for the same parameters as in Fig. \ref{fig:mu03_l0_obsR1}.
The analytic (numerical) ones were obtained using the continued fraction method (Fourier transform of the evolution data).}
\label{tab:num}
\end{table}
\end{center}

A curious behaviour that can be observed in the asymptotically flat case, due to the absence of 
growing modes is displayed in Fig. \ref{fig:mu03_l0_obsR1} (right panel). When the initial data
excites two modes with a similar frequency, one observes a \emph{beating} behaviour. This has
already been noticed in \cite{Witek:2012tr} for an uncharged field in a rotating background and in
\cite{Barranco2013} in a Schwarzschild background. This beating
results from the coupling between the two modes with similar frequency which can be observed in the inset of the figure.

 %
%
%\begin{figure}[h!]
%\begin{center}
%\includegraphics[width=0.37\textwidth,clip,angle=-90]{beating.ps}
%\caption{Evolution of the scalar amplitude  for a Gaussian perturbation with $\phi=3\times 10^{-3}$,  $r_g=10M$, $\sigma =2M$. The
%observation point was $r_0 = 70M$, the mass of the field is $M\mu=0.1$ and
%$\ell =0$, $Mq=0.3$, $Q/M=0.1$. The upper inset is a zoom to show the oscillating behaviour. The Fourier transform is shown
%in the bottom inset and exhibits the two excited modes. 
%}
%\label{fig:beating} 
%\end{center}
%\end{figure}

%
%

%caption {behaviour of the real part $\omega$ of the $\ell= 2$, $n = 0$ QB eigenfrequencies
%as functions of $q$ for different values of $\mu$.}

%In figure, we present some typical examples of the signal when
%$\ell= 0$.

%\newpage

%%%%%%%%%%%%%%%%%%
\subsection{RN BH in a cavity}
%%%%%%%%%%%%%%%%%%

%%%%%%%%%%%%%%%%
%%%%%%%%%%%%%%%
{\bf Gaussian initial data}
%%%%%%%%%%%%%%%
%%%%%%%%%%%%%%%%%

We consider a Gaussian initial data, described by eq. \eqref{gaussian}. Unless otherwise
specified we take $r_m=15M,r_o=5M,\mu M=0.1$.  For some values of the parameters  
the evolution is displayed in Fig. \ref{fig:fourierdom}.  This figure shows that at early times,
$t/M\lesssim 10^4$, the field amplitude (top left panel) decreases; but an exponential growth
follows. The exponential growth becomes clear in the left bottom panel of Fig. \ref{fig:fourierdom},
where the field energy is displayed against a linear fit of the form ${\rm ln}(E/E_{0})$ as a
function of $t/M$ (blue dashed line). In the inset, the initial decrease of the field energy due to
the absorption by the BH is shown. The right panels of Fig.   \ref{fig:fourierdom} show the power
spectrum of the Fourier transform of the field amplitude obtained at four different stages in the
evolution and provide a mode interpretation of the field's development. At stage 1 ($t/M=10^3$) one
identifies three well resolved frequencies.  The dominant mode is the first overtone which has an
exponential decay. The critical frequency, 
separating growing modes ($\omega<\omega_c$) from decreasing modes ($\omega>\omega_c$) is
$\omega_c=0.501431$ and it is marked with a vertical dashed line. At stage 2  ($t/M=10^4$) one
observes a noticeable growth of the fundamental mode which becomes  comparable to the first overtone
at stage 3 ($t/M=4\times 10^4$). Finally, at stage 4 ($t/M=2\times 10^5$), the dominant mode is the
fundamental mode. 
The exponential growth rate for the energy corresponding to the dashed blue line in the bottom left
panel is in
agreement with twice the imaginary part of the frequency of the fundamental mode with an error of less than 1\%.

Next we compare time evolutions for different values of the field charge $q$, keeping the same 
Gaussian initial data as before. We take three cases with $qM=0.8, 1.0, 1.2$, keeping $Q/M=0.9$ and
considering modes with $\ell=1$; the frequencies of the fundamental mode and first two overtones are
displayed in Table \ref{tab:frequencies}. Observe that the unstable modes are the ones with a
positive imaginary part. 

In Fig. \ref{fig:3runsenergy} we display the time evolution of the field amplitude (left panels) 
and power spectrum of the Fourier transform (middle panels) for the three cases described in the
previous paragraph.  In all three cases the field amplitude grows exponentially. The power spectrum,
taken at $t/M=1.5\times 10^5$, shows that the dominant mode is always the one with the largest
(positive) imaginary part in Table \ref{tab:frequencies}, that is, the fastest growing mode. The
right panel of Fig. \ref{fig:3runsenergy} shows an exponential growth for the field energy that
matches twice the imaginary part of the frequency of the corresponding fastest growing mode.

\begin{widetext}
 
\begin{figure}[h!]
\begin{center}
\includegraphics[width=0.35\textwidth,angle=-90]{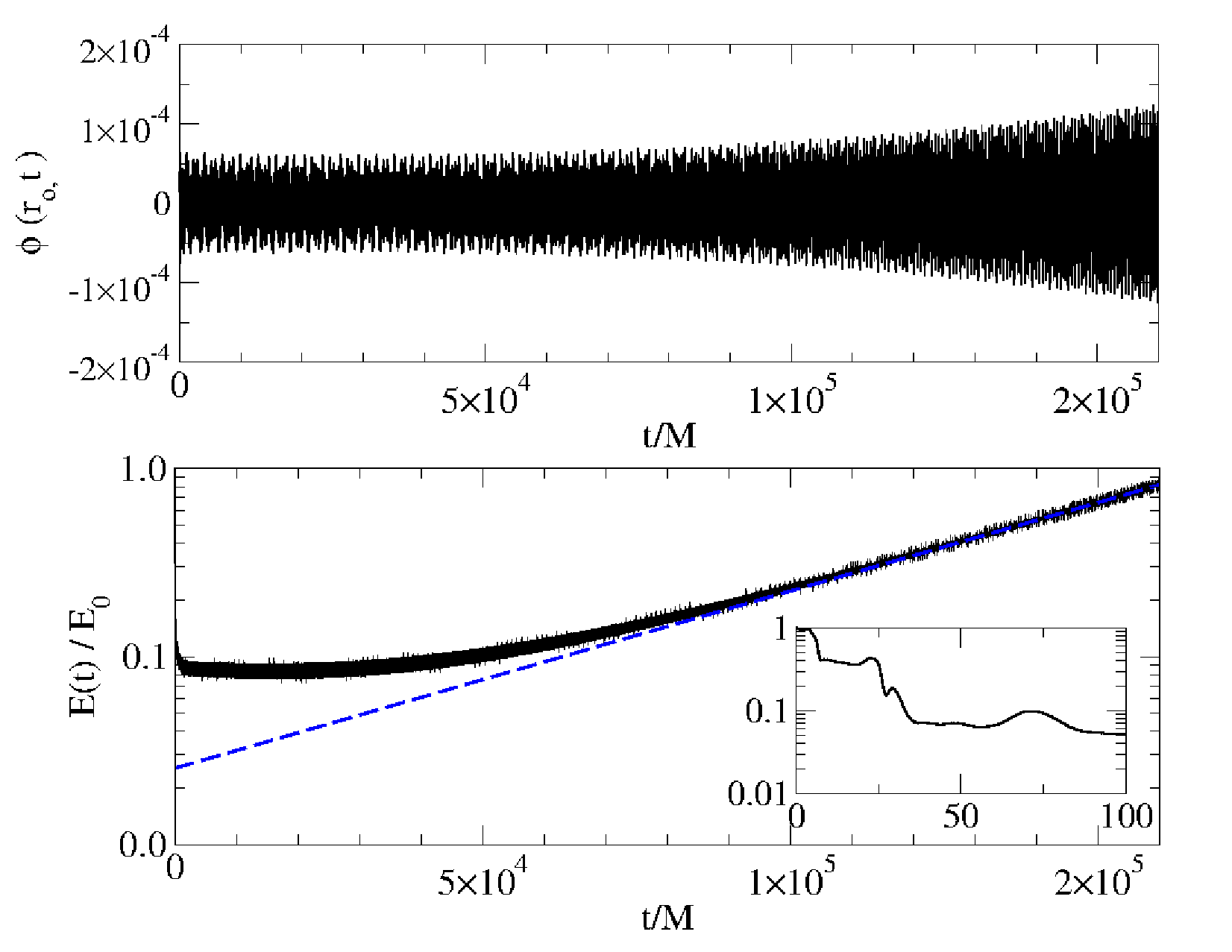}
\includegraphics[width=0.35\textwidth,angle=-90]{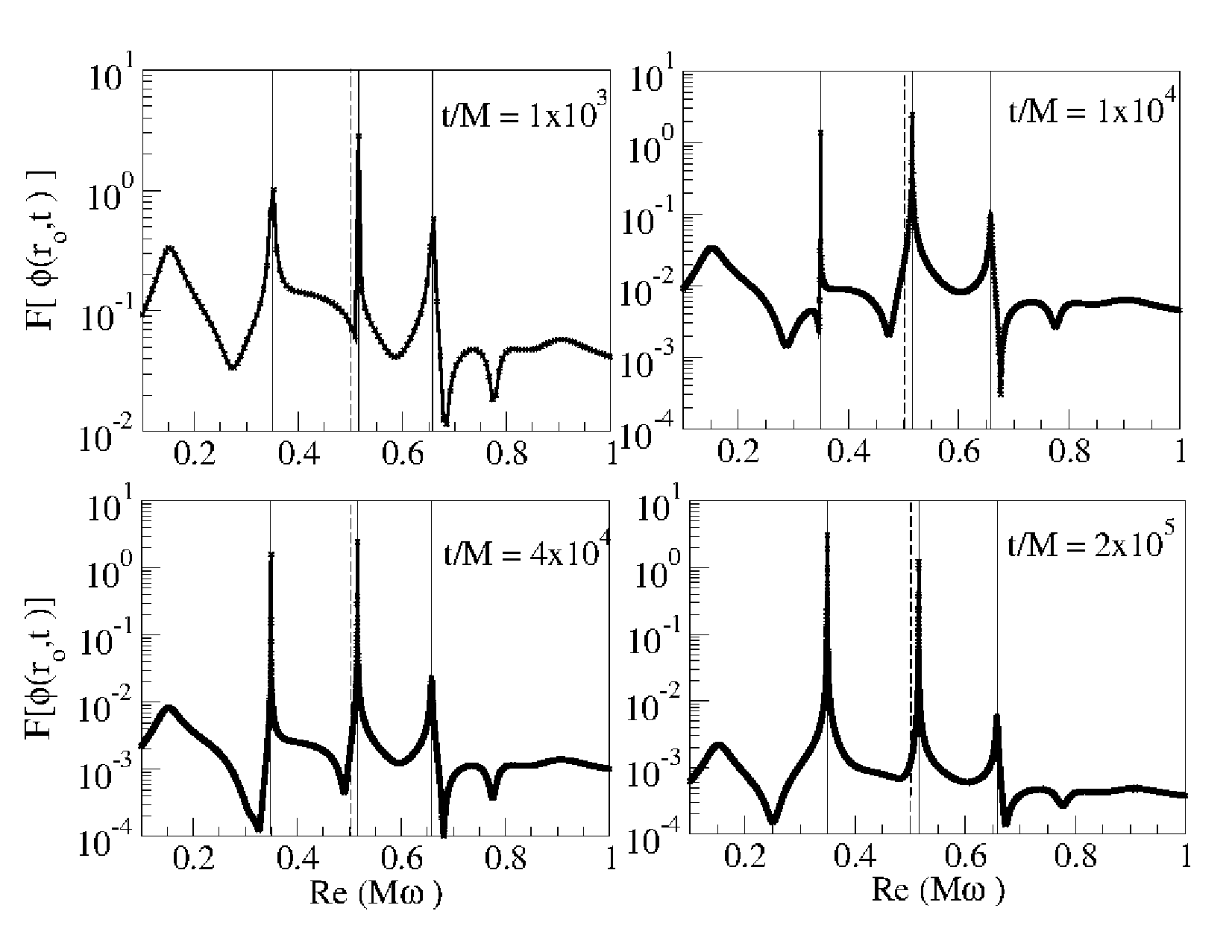}
\caption{
Evolution of a Gaussian wave packet with $r_{cg}=7M$, $\sigma=M$ and $\phi_0=3\times 10^{-4}$ for a
field mode with $\ell=1, qM=0.8$ and a background with $Q/M=0.9$. The top left panel shows the field
amplitude. 
The left bottom panel displays the field energy and the inset details the initial loss of  energy in
the field, due to BH absorption. 
The right panels show the power spectrum of the Fourier transform of the field amplitude at four different time stages.}
\label{fig:fourierdom}
\end{center}
\end{figure}
%\end{widetext}

%\begin{widetext}

\begin{center}
\begin{table}[h!]
  \begin{tabular}{ |c |c | c | c|}
    \hline
$n$ & $Mq=0.8$ & $Mq=1.0$ &$Mq=1.2$   \\ 
\hline \hline
$0$ & $0.349348 + 1.084 \times 10^{-5}i $ &$0.369909 + 2.2807\times 10^{-5} i$ & $0.390017 + 4.300
\times 10^{-5} i$\\
\hline
$1$ & $0.515418 - 7.66693\times 10^{-6} i$ &$0.542963 + 2.7655 \times 10^{-5} i $& $0.569346
+5.469\times 10^{-5} i$ \\
\hline
$2$ & $0.65799 - 1.3986 \times 10^{-3}i$ &$0.694465 - 1.94583\times 10^{-4} i$ & $ 0.727795 +
2.905 \times 10^{-5} i$ \\
\hline
  \end{tabular}
\caption{Complex frequencies for three first mirrored states taking $Q/M=0.9$, $\ell=1$.}
\label{tab:frequencies}
\end{table}
\end{center}
 
%\end{widetext}

%\begin{widetext}

\begin{figure}[h!]
\includegraphics[width=0.34\textwidth,angle=-90]{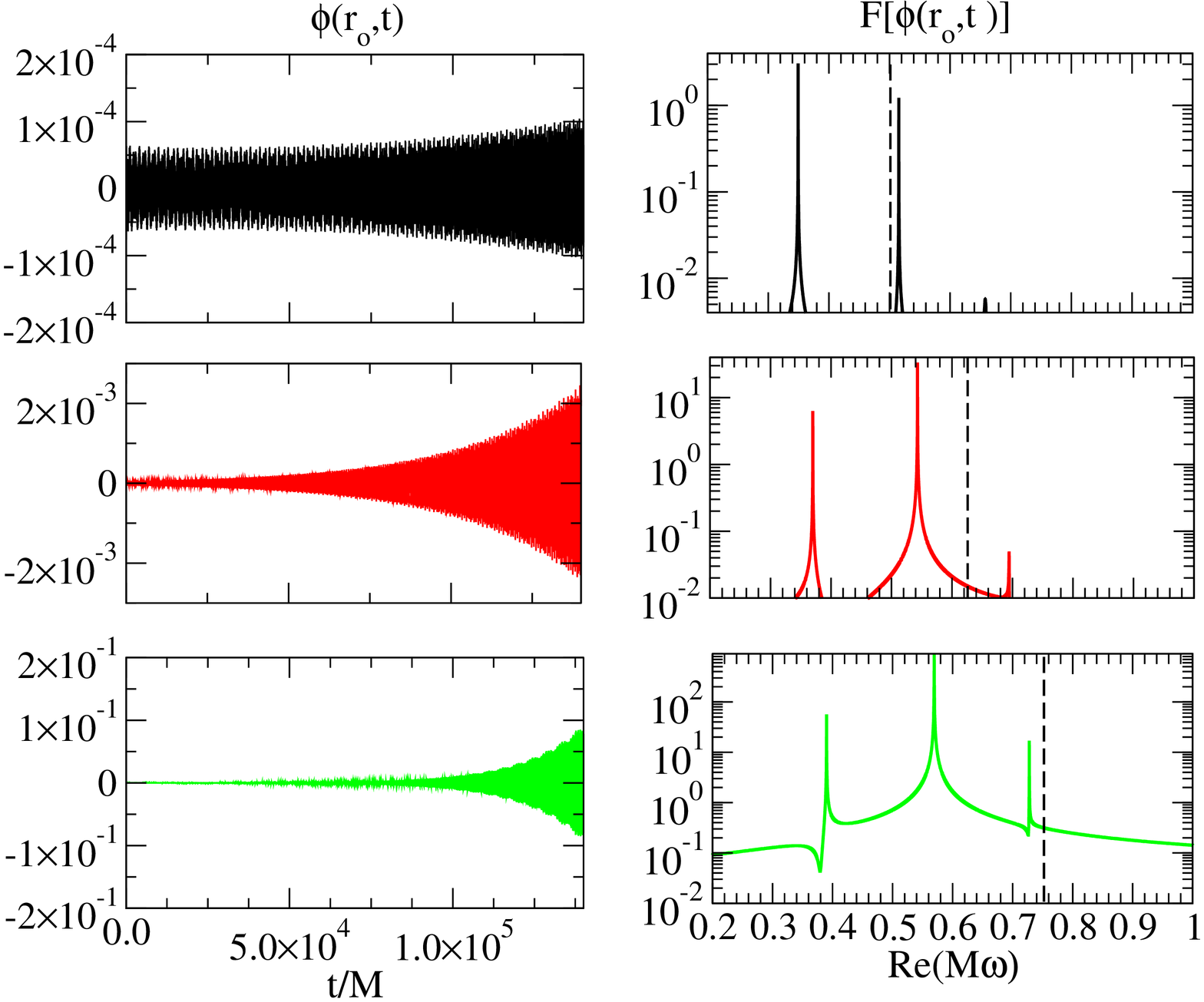}
\includegraphics[width=0.34\textwidth,angle=-90]{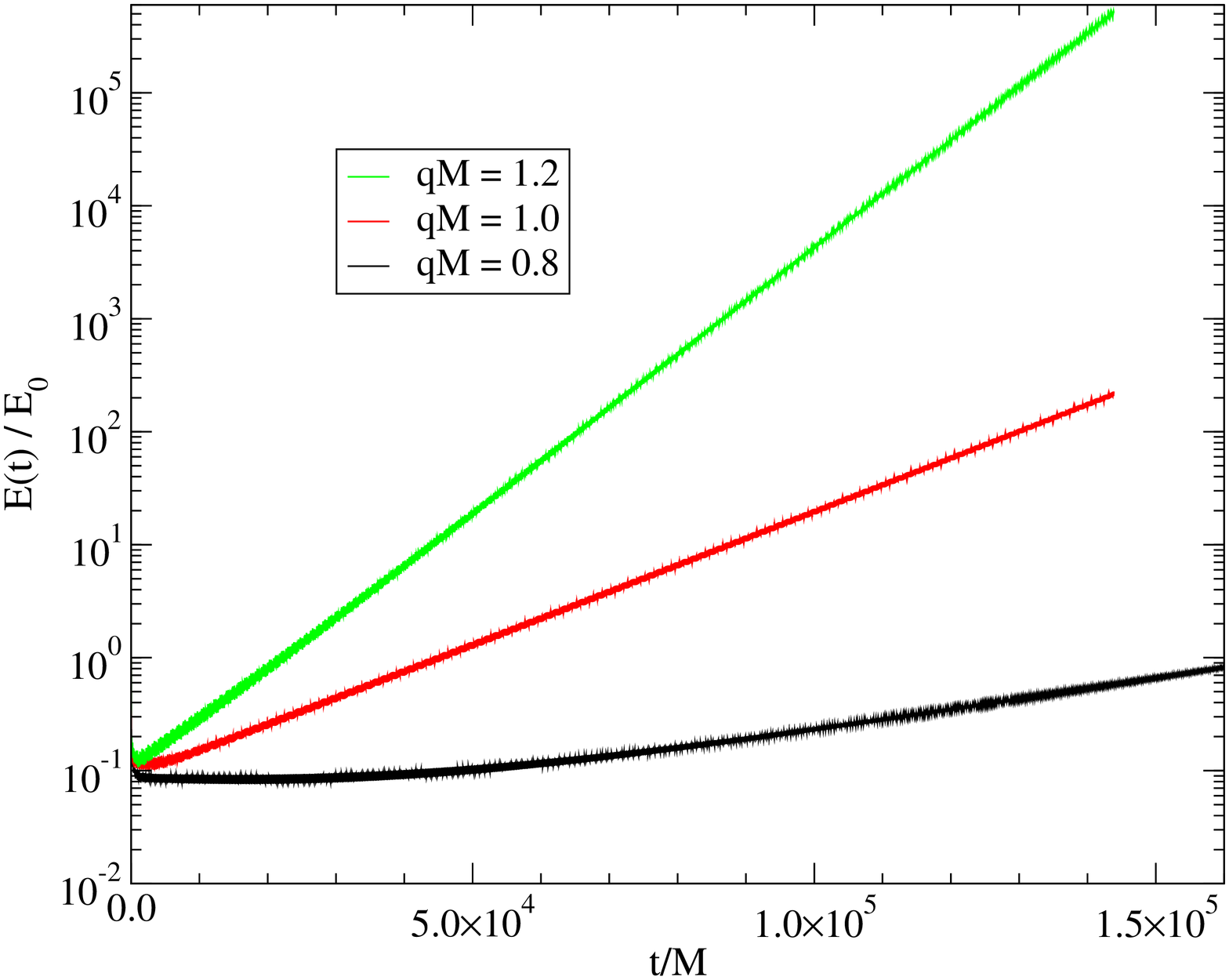}
\caption{(Left panels) Time evolution of the field amplitude for, from top to bottom, 
$Mq=0.8, 1.0, 1.2$, $Q/M=0.9$ and $\ell=1$. (Middle panels) Fourier transform of the times series
data. The dashed lines mark the critical
frequencies $M\omega_c=0.501431$, $M\omega_c=0.626789$,
$M\omega_c=0.752147$.  (Right panel) Normalized energy. 
%The initial values of the energy are $E_0=1.850\times10^{-6}$, $E_0=1.879\times10^{-6}$, and $E_0=1.915\times10^{-6}$ respectively.
}
\label{fig:3runsenergy}
\end{figure}

\end{widetext}

%{\bf Data:}
%\begin{verbatim}
%q=0.8
%0.349348    +0.00001084 I *
%0.515418     -7.66693\times 10^{-6} I
%0.65799     -0.0013986 I

%q=1.0
%0.369909   +0.000022807 I
%0.542963   +0.000027655 I * 
%0.694465   -0.000194583 I

%q=1.2
%0.390017   +0.00004300 I 
%0.569346   +0.00005469 I *
%0.727795   +0.00002905 I
%0.866873   -0.00137116 I 
%\end{verbatim}

We also consider the evolution of $\ell=0$ modes to show they exhibit the superradiant 
instability and compare their growth time rates with the $\ell=1$ modes. Superradiant $\ell=0$ modes
are particular to charged BHs, and do not occur in the rotating case. For the same $q,Q$ parameters
as before we find that the frequencies of the fundamental mode and first two overtones are those
displayed in Table \ref{tab:frequenciesl0}. Then, in Fig. \ref{fig:3runsenergyl0} the time evolution
of these modes is displayed in analogy to Fig. \ref{fig:3runsenergy}. A noticeable fact is that the
growth rates are about two orders of magnitude faster than than the corresponding $\ell=1$ modes.

%\bigskip

%\newpage

\begin{widetext}

\begin{center}
\begin{table}[h!]
  \begin{tabular}{ |c |c | c | c|}
    \hline
$n$ & $Mq=0.8$ & $Mq=1.0$ &$Mq=1.2$   \\ 
\hline \hline
$0$ & $0.28936  + 2.10503 \times 10^{-3}i $ &$0.31285 + 2.78491\times 10^{-3} i$ &
$0.33541  + 3.43661 \times 10^{-3} i$\\
\hline
$1$ & $0.44594  +0.63712\times 10^{-3} i$ &$ 0.47943 +2.75014 \times 10^{-3} i $&
$0.51008 + 3.44575 \times 10^{-3} i$ \\
\hline
$2$ & $0.57591 -8.47143 \times 10^{-3}i$ &$0.62207 +2.80632\times 10^{-4} i$ & $
0.66325 + 2.83061 \times 10^{-3} i$ \\
\hline
  \end{tabular}
\caption{Complex frequencies for the three first mirrored states taking $Q/M=0.9$, $\ell=0$.}
\label{tab:frequenciesl0}
\end{table}
\end{center}
 
%\end{widetext}
%\begin{widetext}

\begin{figure}[h!]
\includegraphics[width=0.34\textwidth,angle=-90]{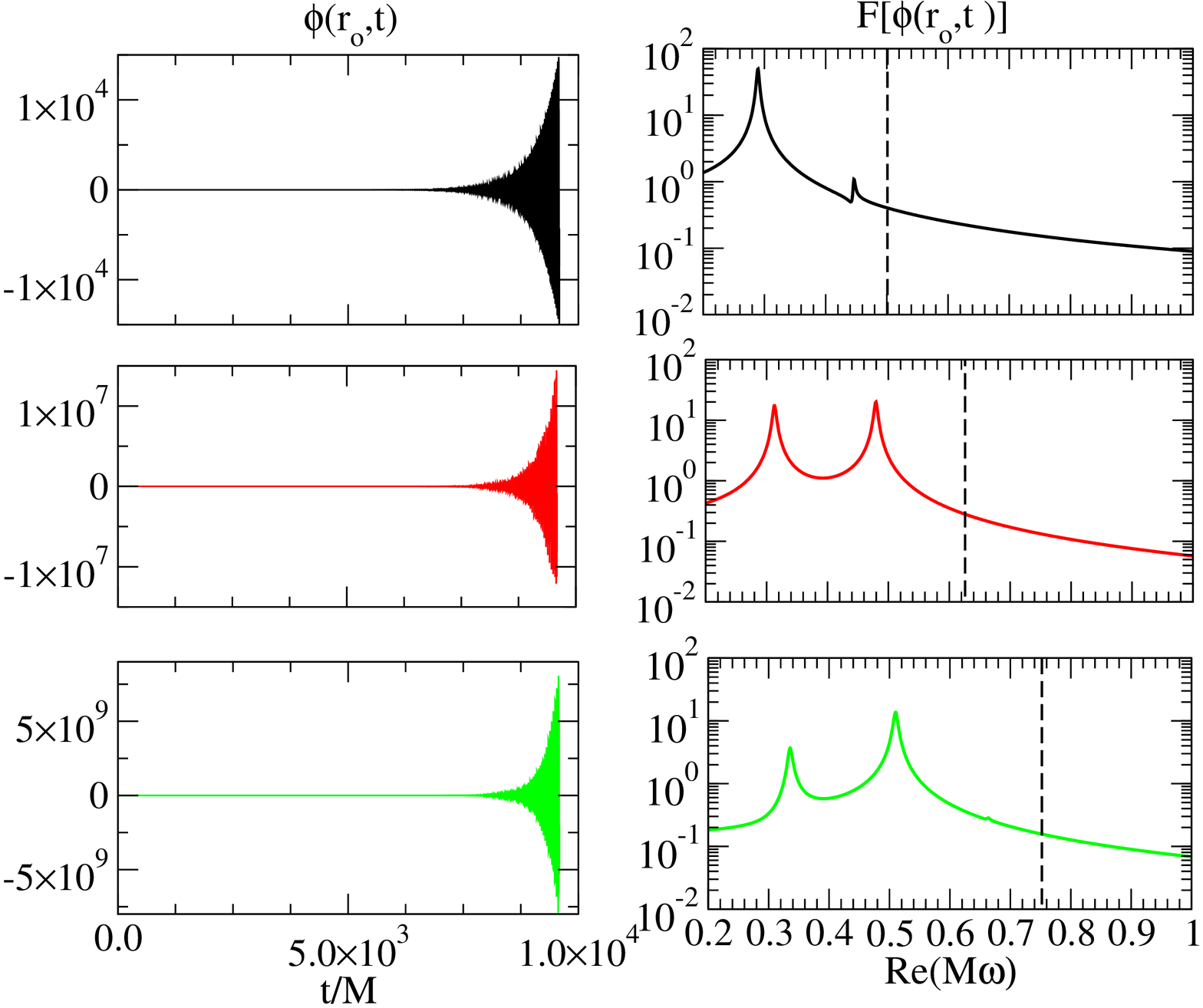}
\includegraphics[width=0.34\textwidth,angle=-90]{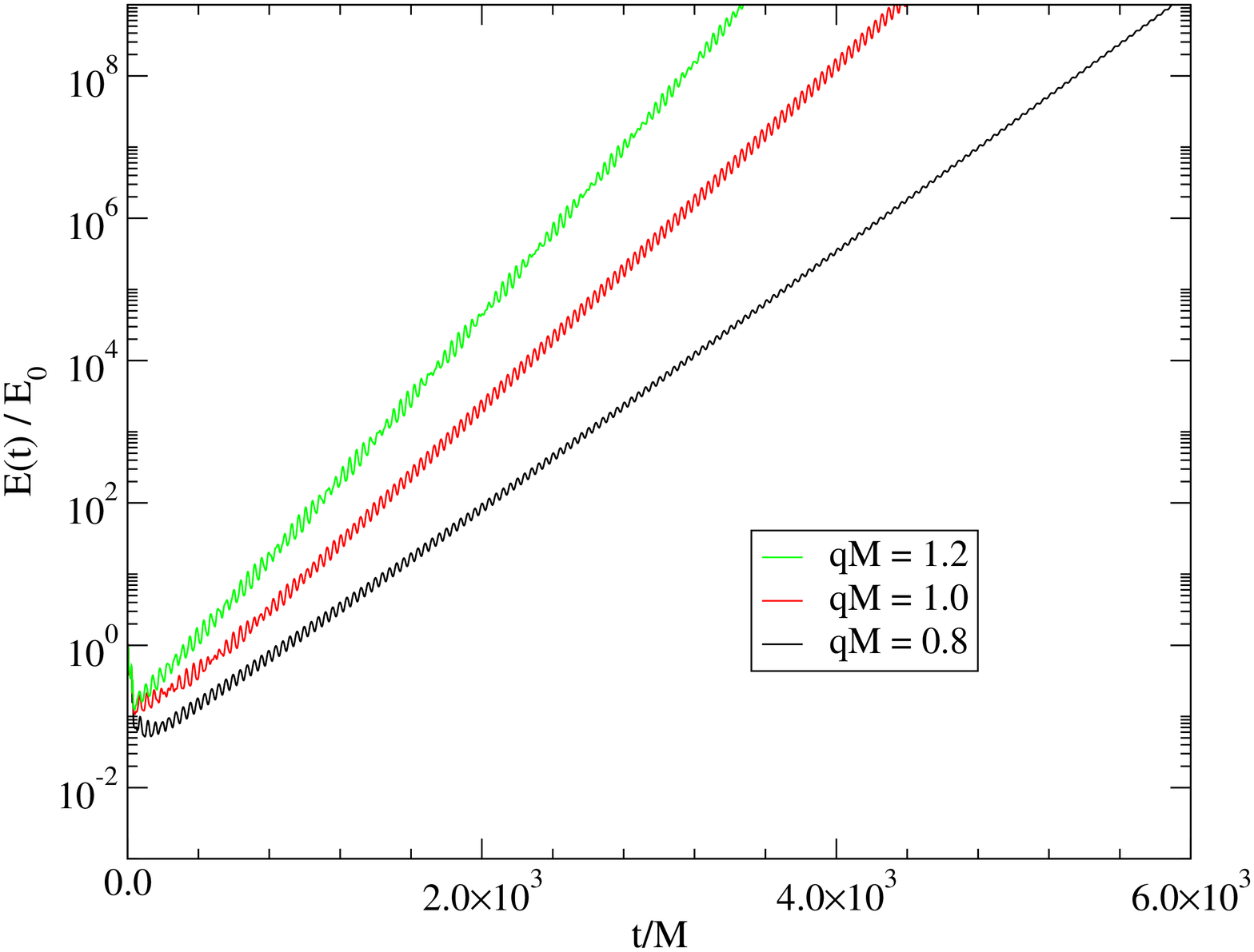}
\caption{(Left panels) Time evolution of the field amplitude for, from top to bottom, 
$Mq=0.8, 1.0, 1.2$, $Q/M=0.9$ and $\ell=0$. (Middle panels) Fourier transform of the times series
data. The dashed lines mark the critical
frequencies $M\omega_c=0.501431$, $M\omega_c=0.626789$,
$M\omega_c=0.752147$. Note that critical frequencies are $\ell$-independent. (Right panel) 
Normalized energy.}
\label{fig:3runsenergyl0}
\end{figure}
\end{widetext}

\bigskip

%%%%%%%%%%%%%%
{\bf Regularized mirrored states}
%%%%%%%%%%%%%%

\begin{widetext}

\begin{figure}[h!]
\begin{center}
\includegraphics[width=0.35\textwidth,angle=-90]{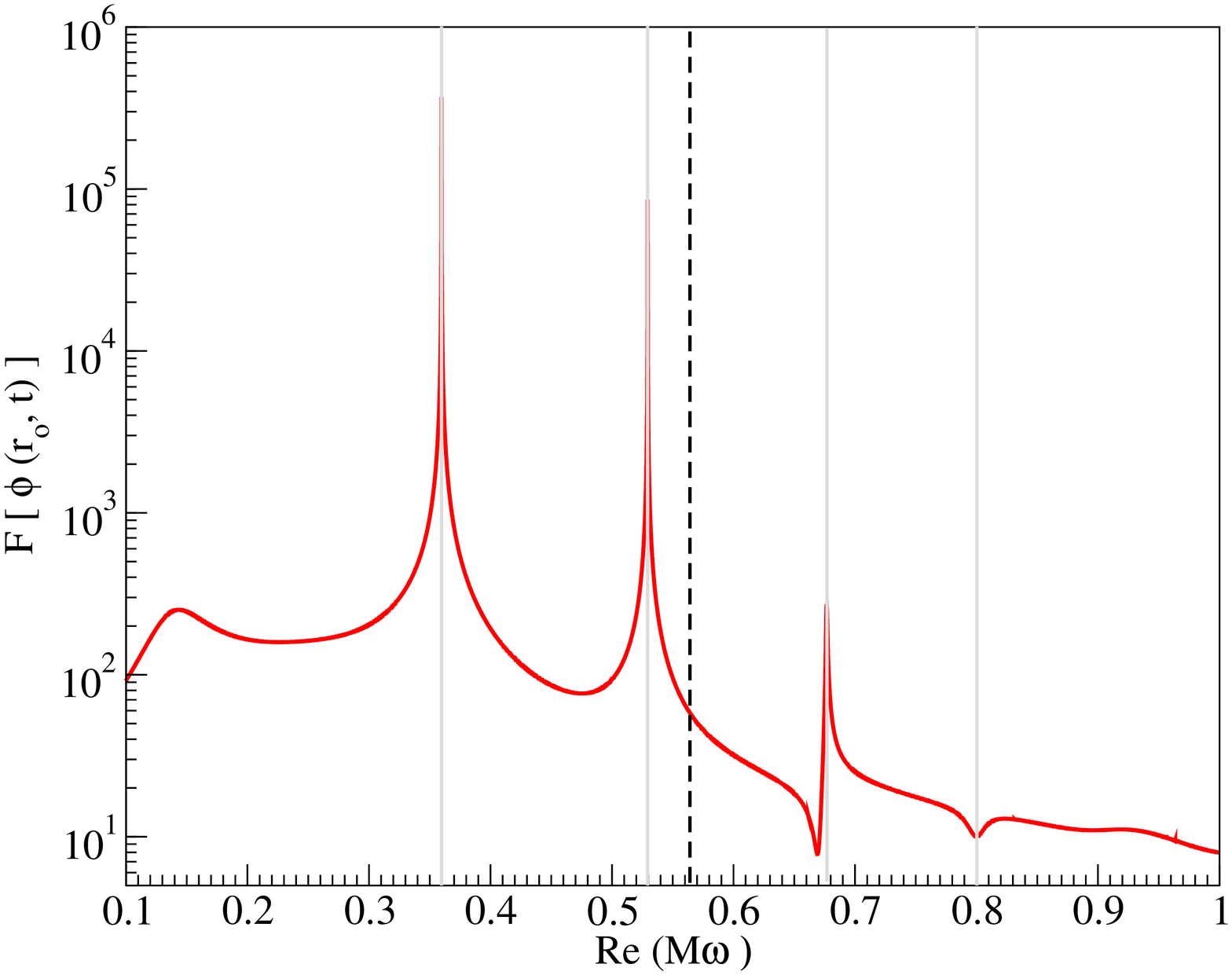}
\includegraphics[width=0.35\textwidth,angle=-90]{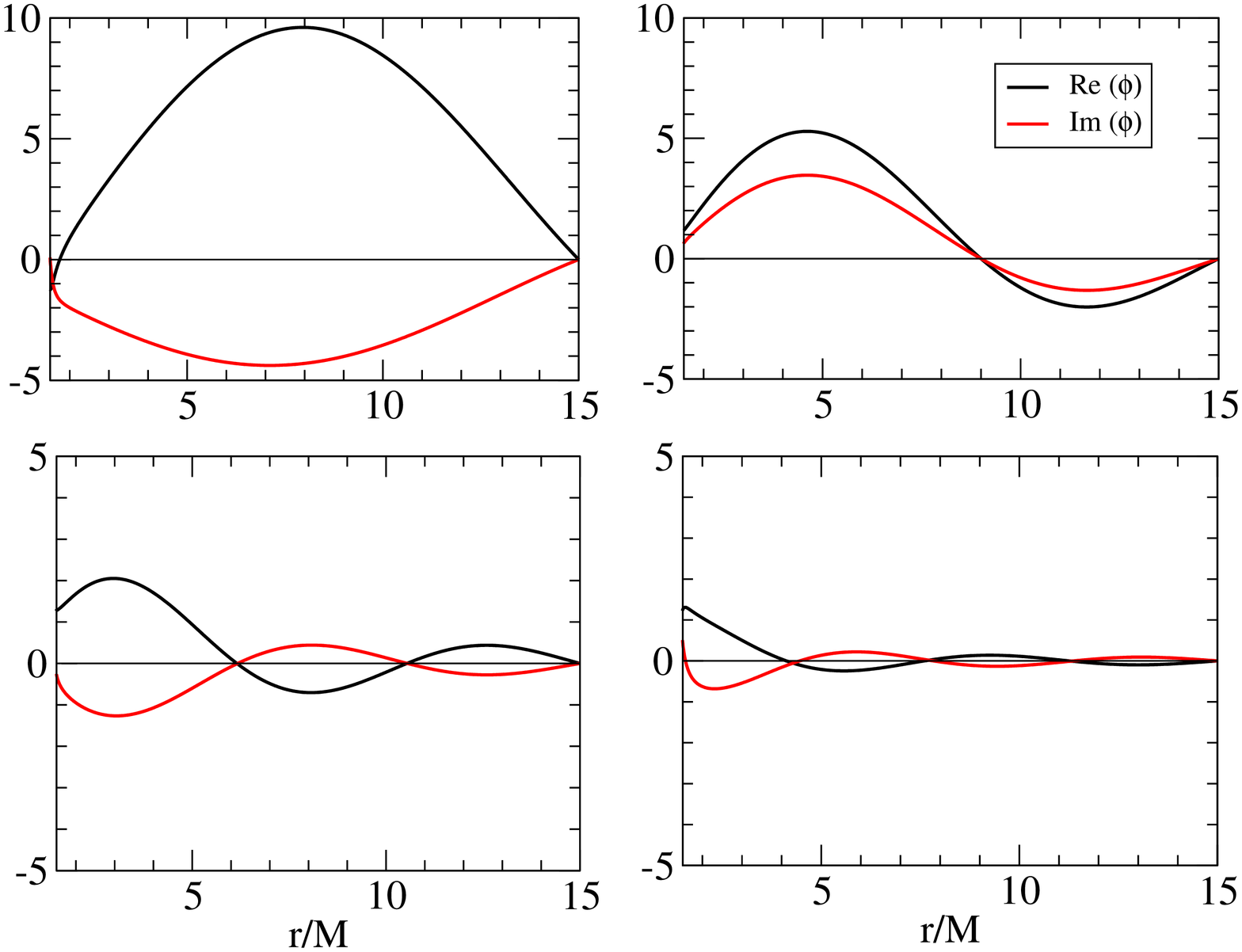}
\caption{(Left panel) Fourier transform of the field amplitude for a regularized mirrored state: the fundamental mode for 
$\ell=1, Q/M=0.9, qM=0.9$. The critical frequency is $M\omega_c= 0.56411$ (dashed line). The vertical lines denote the first overtones
obtained with direct integration of the radial equation. (Right panel) Real and imaginary part of the
 ground state of the field and the first 3 overtones.
}
\label{fig:Am_l1_Q09_q09freq}
\end{center}
\end{figure}
\end{widetext}

The previous examples with Gaussian initial data, as well as other numerical experiments, show 
that for a given set of parameters $q,\mu,Q,\ell, r_m$, the corresponding fundamental mode and the
first overtones are excited. Perhaps all overtones are excited, but for higher overtones, which are
energetically more costly,  the corresponding amplitude is too small to be seen in our numerical
evolutions. One may ask, however, if it is possible to choose initial data as to excite a
\textit{single} frequency. The obvious initial data to attempt such state, is to take the radial
function of the mirrored sate. These radial functions obey eq. \eqref{eq:Slike} and may be
constructed in the frequency domain \cite{Herdeiro:2013pia}. We have consider the time evolution of
such an initial state: the fundamental mode for $qM=0.9=Q/M$, $\ell=1$. The Fourier transform of the
time evolution of the field's amplitude is shown in  Fig. \ref{fig:Am_l1_Q09_q09freq}. We observe
that higher overtones appear excited as well. The 
reason may 
be related to the fact that mirrored states (as in the generic case of quasi-bound states) 
diverge at the event horizon and the radial function we have used to mimic the quasi-bound state is
smoothed out therein. Although the two radial functions only differ in a small vicinity of the
horizon this may be enough to excite other states.

%\begin{figure}
%\begin{center}
%\includegraphics[width=0.4\textwidth,angle=-90]{Radmodes.ps}
%\caption{The real and imaginary part of the field, ground state with 3 overtones. }
%\label{fig:Radmodes}
%\end{center}
%\end{figure}

%%%%%%%%%%%%%%%%%%
{\bf Rapid growth configurations}
%%%%%%%%%%%%%%%%%%

\begin{figure}
\begin{center}
\includegraphics[width=0.4\textwidth,angle=-90]{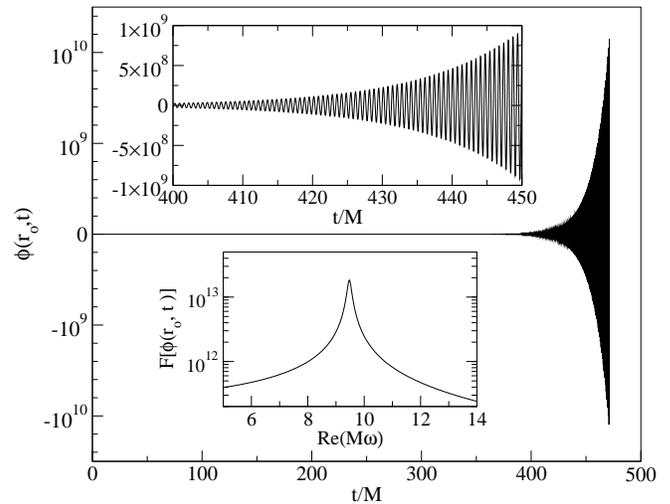}
\caption{Rapid growing mode. The parameters are $\ell=1$, $Q/M=0.9$, $qM=40$. The main panel shows the signal amplitude. 
The upper inset is a zoom to show the oscillating behaviour. The Fourier transform is shown
in the bottom inset.}
\label{fig:extrem_q40}
\end{center}
\end{figure}

Finally we shall consider the fastest growing configurations discussed in 
\cite{Herdeiro:2013pia}. We thus take $r_m=5M$, $qM=40$,
$Q/M=0.9$ and $\ell=1$ for which a mode was reported with $M\omega=9.4655 + 0.07099i$. 
For the same parameters the most unstable mode with $\ell=0$ has $M\omega=9.4385 + 0.08157i$.
Thus for this values of the charge the difference in growth rate for $\ell=0$ and $\ell=1$ modes has
already become very small. We checked this is the general pattern: as $q$ increases the dependence
in $\ell$ becomes less important in agreement with the results of \cite{Hod:2013fvl}.  Fig.
\ref{fig:extrem_q40} shows the signal as measured by an observer located at $r_o=3M$. 
As in the previous cases we took a Fourier transform of the signal to obtain the oscillating
frequency.
In this high frequency set up the dominant mode is the ground state. At very early times one can see
the growth of the signal. Although we do not have enough data to say whether other overtones are
present (longer runs will drive the energy very outside of the test field regime) the frequency
agrees with the value previously reported. The growth rate coincides with the imaginary part of
the frequency as well.

%{\bf frequencies:}
%\begin{verbatim}
%
%9.4655     +0.0712526      
%10.7166    +0.0503      
%
%\end{verbatim}

%%%%%%%%%%%%%%%%%%%%%%%
\section{Conclusions}
\label{conclusions}
%%%%%%%%%%%%%%%%%%%%%%%
The laws of BH mechanics do not forbid physical processes that may extract energy 
from BHs. Indeed, for a Kerr-Newman BH, the first and second laws require only that the BH energy
variation $dE$ is bounded below by the angular momentum and charge variations,  $dJ,dQ$
respectively, as
\[
dE\ge \Omega_HdJ+\Phi_HdQ \ .
\]
Thus energy extraction, i.e $dE<0$, is possible if accompanied by a sufficient  large extraction 
of either angular momentum or charge. A physical process that materializes this possibility is
superradiant scattering. Taking $dE=-\omega$, $dJ=-m$, $dQ=-q$ one obtains precisely the
superradiance conditions described in the Introduction.

The existence of quasi-bound states of a bosonic field in the superradiant regime generates an 
instability of the BH. The field extracts energy from the BH, together with angular momentum and
charge (or both), and starts to pile up around it until non-linear phenomena start to dominate the
evolution. Understanding such evolution requires non-linear numerical simulations. Some progress
in this type of simulations for the Kerr case has been reported in 
\cite{Yoshino:2012kn,Yoshino:2013ofa}, where non-linear interactions for the scalar field were
considered but not the backreaction on the geometry. The fully-non linear system for the coupled
Einstein-scalar field equations seem quite challenging for this case. Even though the numerical
relativity community has available techniques to tackle such problems \cite{Cardoso:2012qm}, the
very slow growth rates require very long time evolutions, and above all, preventing numerical errors
from masking the physical signal.

In this paper we have reported linear time evolutions for a charged scalar field in the 
background of a RN BH and in particular observed the superradiant growth - in the linear regime - of
the field when the BH is enclosed in a cavity. Our results further confirm the existence of rapid
growing unstable modes - as compared to the Kerr case - and which may be spherically symmetric 
\cite{Herdeiro:2013pia}.  These two features provide a considerable advantage in performing fully
non-linear numerical simulations. Such simulations will allow monitoring the development and understanding the end-point of the superradiant
instability in a setup which, albeit not exactly equal, will share common features to the
astrophysically more relevant Kerr case. 

%Future work may involve a self gravitating scalar field which allows for the 
%back-reaction of the spacetime. 
%This type of studies will shed some light about the end state of
%the superradiant instability in a more realistic scenario.

%The system scalar field black hole with a mirror can be regarded, in the test field limit, as the
%evolution of a scalar field with a step potential $V(r)=V_0\Theta (r-r_0)$ where $\Theta(r)$ is the
%Heaviside function and $r_{0}$ is the mirror position. However, one can mimic the mirror like
%behaviour if one imposes that the field vanishes at the mirror position. 

%%%%%%%%%%%%%%%%%%%%%%%%%%%%%%%%%%%%%%%%%%%%%%%
\section*{Acknowledgements}
%%%%%%%%%%%%%%%%%%%%%%%%%%%%%%%%%%%%%%%%%%%%%%%

J. C. D. acknowledges support from CONACyT-M\'exico and from FCT via project No. 
\textit{PTDC/FIS/116625/2010}. This work was also supported by the \textit{NRHEPÐ295189
FP7-PEOPLE-2011-IRSES} Grant.
%%%%%%%%%%%%%%%%%%%%%%
%%%   REFERENCES   %%%
%%%%%%%%%%%%%%%%%%%%%%

\bibliographystyle{h-physrev4}
\bibliography{num-rel}

%%%%%%%%%%%%%%%
%%%   END   %%%
%%%%%%%%%%%%%%%
\end{document}